\documentclass{JINST}
\let\ifpdf\relax
\pdfoutput=1

\newcommand{\ckcs}{~keV$^{-1}$cm$^{-2}$s$^{-1}$ }

\title{X-ray detection with Micromegas with background levels below 10$^{-6}$\ckcs }

\author{ S.~Aune$^a$, F.~Aznar$^b$\footnote{Present addr.: Centro Universitario de la Defensa, Universidad de Zaragoza, Ctra. de Huesca s/n, 50090 Zaragoza, Spain} , D.~Calvet$^a$, T.~Dafni$^b$, A.~Diago$^b$, F.~Druillole$^a$ , G.~Fanourakis$^c$, E.~Ferrer-Ribas$^a$, J.~Gal\'an$^b$, J.A.~Garc\'ia$^b$, A.~Gardikiotis$^d$, J.G.~Garza$^b$\footnote{Corresponding author}, T.~Geralis$^d$, I.~Giomataris$^a$, H.~G\'omez$^b$, D.~Gonz\'alez-D\'iaz$^b$,  D.C.~Herrera$^b$, F.J.~Iguaz$^{a,b}$, I.G.~Irastorza$^b$, D.~Jourde$^a$, G.~Luz\'on$^b$, H.~Mirallas$^b$, J.P.~Mols$^a$, T.~Papaevangelou$^a$, A.~Rodr\'iguez$^b$, L.~Segu\'i$^{b}$\footnote{Present addr.: Physics Department, University of Oxford, The Denys Wilkinson Building, Keble Road, Oxford,
OX1 3RH, UK
}, A.~Tom\'as$^{b}$\footnote{Present addr.: Brackett Laboratory, Imperial College, London, UK}, T.~Vafeiadis$^e$, S.C.~Yildiz$^{f,g}$  \\
\llap{$^a$} Centre d'\'Etudes de Saclay, CEA, Gif-sur-Yvette, France\\
\llap{$^b$} Grupo de F\'isica Nuclear y Astropart\'iculas, University of Zaragoza, Zaragoza, Spain\\
\llap{$^c$} Institute of Nuclear Physics, NCSR Demokritos, Athens, Greece\\
\llap{$^d$} University of Patras, Patras, Greece\\
\llap{$^e$} CERN, European Organization for Particle Physics and Nuclear Research, Geneva, Switzerland\\
\llap{$^f$} Do\u{g}u\c{s} University, Istanbul, Turkey\\
\llap{$^g$} Bo\u{g}azi\c{c}i University,  Istanbul, Turkey\\
  E-mail: \email{jgraciag@cern.ch}}

\abstract{Micromegas detectors are an optimum technological choice for the detection of low energy x-rays. The low background techniques applied to these detectors yielded remarkable background reductions over the years, being the CAST experiment beneficiary of these developments. In this document we report on the latest upgrades towards further background reductions and better understanding of the detectors' response. The upgrades encompass the readout electronics, a new detector design and the implementation of a more efficient cosmic muon veto system. Background levels below 10$^{-6}$\ckcs have been obtained at sea level for the first time, demonstrating the feasibility of the expectations posed by IAXO, the next generation axion helioscope. Some results obtained with a set of measurements conducted in the x-ray beam of the CAST Detector Laboratory will be also presented and discussed. 
}


\begin{document}

%section: INTRO
\section{Introduction}\label{sec:intro}
Low-mass and weakly interacting pseudoescalar particles are invoked in many extensions of the Standard Model, including string theory ~\cite{Ringwald:2012cu}. One prominent example of these particles is the axion, which arises in the most widely accepted solution to the strong CP problem ~\cite{Peccei:1977hh}. The axion is naturally a very compelling candidate for constituting partially or totally the dark matter of the Universe ~\cite{Abbott:1982af,Dine:1982ah,Preskill:1982cy}. 

The detection of the axion could be within the reach of current or near future experiments. Axion helioscopes are an important category of experimental searches, which rely on the back-conversion of axions generated in the Sun into x-rays by means of a laboratory magnetic field. The CERN Axion Solar Telescope (CAST) ~\cite{Zioutas:2004hi} is the most powerful implementation of a helioscope, setting leading limits in the axion-photon coupling for a wide range of axion masses. Two of the parameters driving the sensitivity of the experiment are the x-ray detection efficiency and the background level of the x-ray detectors. Three of the four CAST magnet bores are equipped with TPCs based on microbulk Micromegas readouts~\cite{Giomataris:1995fq, Andriamonje:2010zz, Iguaz:2011xi}, conceived to detect the converted photons in the 2--7 keV CAST Region of Interest (RoI). Two of the Micromegas detectors can operate in axion-sensitive mode (i.e. magnet pointing to the Sun) during the sunset and the other during the sunrise. Since the expected number of signal events is very low, the background level of the detectors must be reduced as much as possible. The discovery potential of an axion helioscope can be expressed in terms of a figure of merit, $f$, which it is inversely proportional to the minimal signal stregth to which the experiment is sensitive to, g$_{a\gamma}^{-4}$. The contribution to $f$ of the detectors and optics is 

\begin{equation}
f_{DO}=\frac{\epsilon_d \epsilon_o}{\sqrt{b a}}
\end{equation}

where $\epsilon_d$ is the detectors' efficiency, $\epsilon_o$ is the focusing efficiency, $b$ is the normalized background level of the detector in area and time and $a$ is the focusing area \cite{Irastorza:2011gs}.
    There are two complementary strategies to increase $f_{DO}$: signal focalization and low background detectors. The first strategy is exemplified by the CCD+ABRIXAS telescope detection line in CAST~\cite{kuster2007}, while the second one by the other three detection lines based on low background Micromegas detectors. For the forthcoming data taking campaign there are plans to merge both strategies. This fact will increase the signal-to-noise ratio (S/N ratio) of the Micromegas detector. Additionally, it will also serve as a pathfinder project to test the technological options being proposed to build large scale, cost effective, x-ray optics with customized parameters for the recently proposed International Axion Observatory (IAXO)~\cite{Irastorza:2011gs,Irastorza:1567109}, the future axion helioscope. IAXO requirements highly motivate the development of low background x-ray detection strategies. The goal background level of IAXO is 10$^{-7}$--10$^{-8}$ \ckcs, while in CAST-2013, Micromegas detectors have already demostrated levels as low as 6$\times$10$^{-7}$ \ckcs.
 
The implementation of these strategies in CAST over the years, many of them imported from deep underground experiments (e.g. shielding design, radiopurity), have provided a background reduction of two orders of magnitude (see Fig. \ref{fig:fig10}).  The applied techniques that led to such reductions are explained in detail in a recent paper~\cite{Aune:2013pna}. While the reader is referred to that document for a complete description of the CAST-Micromegas, a background model of the current detectors and an explanation of low background strategies, this document focuses on the very latest upgrades, tests and results of the CAST 2013 data taking campaign.

The last upgrades concern the readout electronics, presented in Section \ref{sec:electronics}. The new detector design and the characterization of the first set of them is presented in Section \ref{sec:character}. The effect on the CAST-Micromegas background level of the implementation of a dedicated custom-made cosmic muon veto is presented in Section \ref{sec:veto}. Finally, the characterization of CAST-Micromegas detectors using  Particle Induced X-ray Emission (PIXE) to produce mono-energetic x-ray beams is presented in Section \ref{sec:xraybeam}. The current lines of research towards lower backgrounds are described in Section \ref{sec:proscpects}, where we show the feasibility of fulfilling IAXO requirements.

%section: ELECTRONICS Upgrade
\section{CAST-Micromegas electronics upgrade}\label{sec:electronics}

\begin{figure}[t]
\centering
\includegraphics[height=0.27\textwidth]{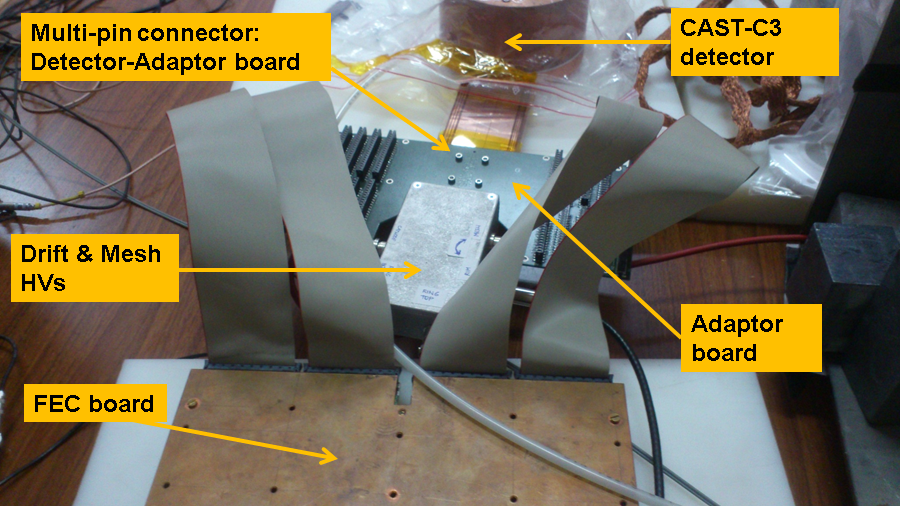} 
\includegraphics[height=0.27\textwidth]{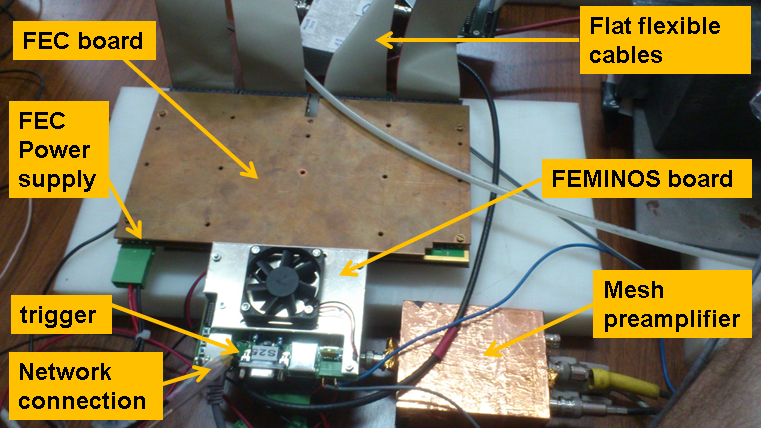}
\caption{CAST-C3 Micromegas detector read out by AFTER-based electronics. The gray flat cables bring the strips' signals from the detector connector to the FEC (shielded in the picture with a copper plate) using a custom-made adaptor card. The HVs are supplied via the same printed circuit glued on the copper base. The signal induced on the mesh electrode after being preamplified is used for triggering the electronics. Data is gathered by the DAQ system via a standard network connection.}
\label{fig:fig0}
\end{figure}

\begin{figure}[b]
\centering
\includegraphics[height=0.27\textwidth]{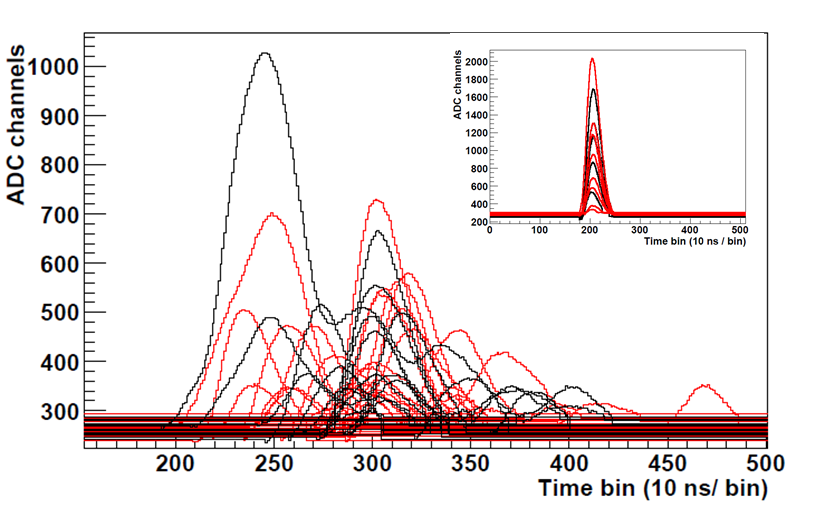}
\includegraphics[height=0.27\textwidth]{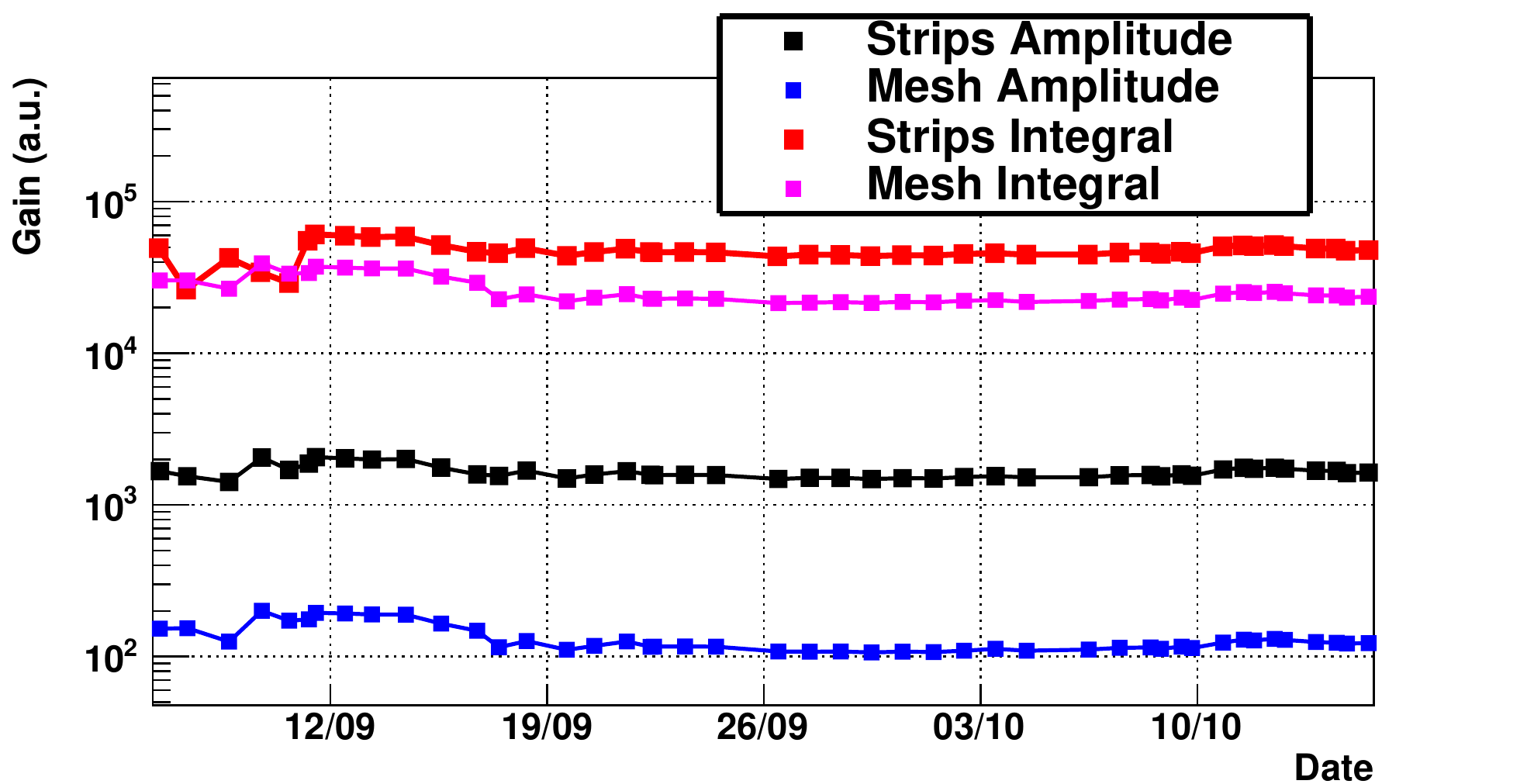}
\caption{Left: $x$ (black) and $y$ (red) strip pulses of CAST-M10 detector for a background event. The embedded figures show the pulses of an x-ray calibration event for comparison. Right: gain evolution of the CAST-M18 detector in operation at CAST during the 2013 data taking campaign.}
\label{fig:fig1}
\end{figure}

 The readout electronics of the CAST-Micromegas detectors have been upgraded for the 2013 data taking campaign. The former Gassiplex-based electronics \cite{Santiard:1994ps} have been replaced by the AFTER chip based electronics developed for the readout of the large T2K time projection chambers ~\cite{Baron:2008zza,Baron:2010zz}. This readout system comprises modular fast electronics that amplify and digitize the pulse signal of every strip (see Fig. \ref{fig:fig1}, left), contrarily to the Gassiplex electronics that just provide the integrated value of the charge signal of each strip. The main component of this readout electronics is the 72-channel AFTER (ASIC For TPC Electronics Readout) chip. It is a very versatile electronics as several parameters are adjustable (sampling rate, shaping time, gain, test mode, etc.), allowing the optimization of the offline analysis. The Front End Card (FEC) comprises four of these chips, enabling to read 288 channels with a single board. CAST-Micromegas detectors have 106--120 strips in each axis, so a single FEC is enough to read out the whole detector. An adaptor board was designed and built to bring the CAST-Micromegas physical channels to the FEC by means of four ERNI-ERNI flat flexible cables (see Fig. \ref{fig:fig0}). The AFTER ASIC has not autotrigger capabilities so an external trigger is generated from the amplified signal of the mesh, which is sampled at 1 GHz frequency in a 2.5 $\mu$s window by a 12-bit dynamic range VME digitizing board, the MATACQ (MATrix for ACQuisition)\cite{MATACQ}. The AFTER ASIC collects and samples the detector signals continuously at 100 MHz in 511 samples per channel, recording a window of $\sim$5 $\mu$s, which is longer than the maximum drift time of the CAST-Micromegas detectors.  When the readout electronics is triggered by the mesh pulse, the analog data from all the channels are passed to an ADC converter. A pure digital electronics card, the FEMINOS board~\cite{DCalvet:Feminos},  gathers the data and performs pedestal subtraction. Finally, those samples whose value is above a user-defined number of standard deviations from the pedestals mean for each channel are sent to the DAQ system by means of a standard network connection, performing therefore an important data reduction.

\begin{figure}[htb!]
\centering
\includegraphics[height=0.28\textwidth]{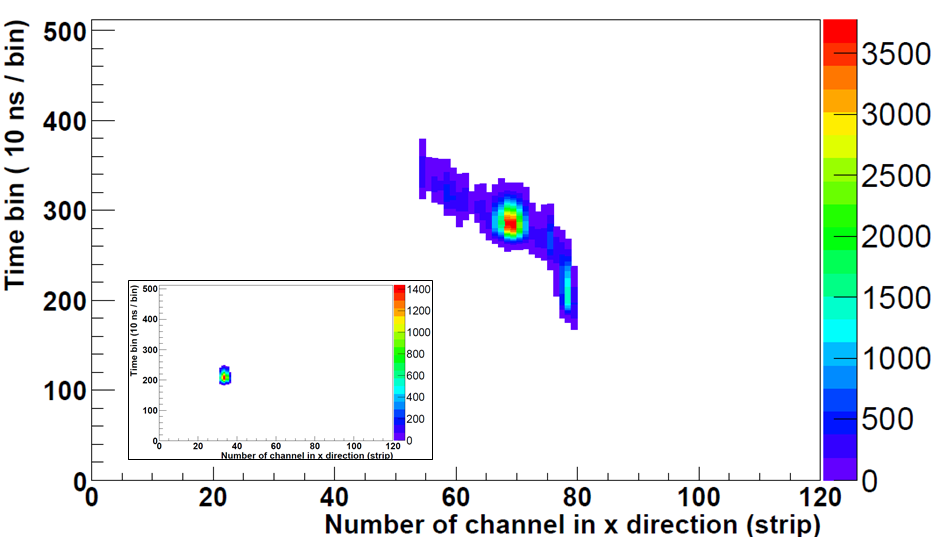} \hspace{5mm}
\includegraphics[height=0.28\textwidth]{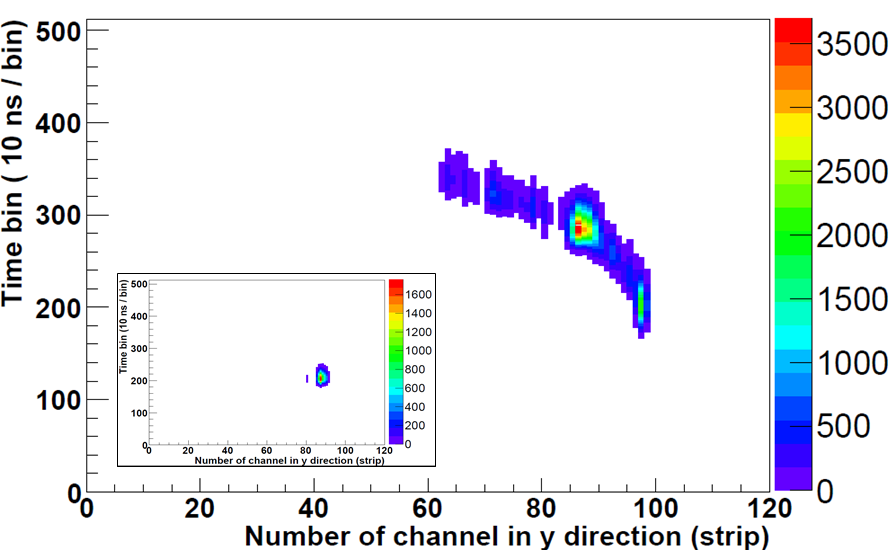}
\caption{The $xz$ (left) and $yz$ (right) view  of an electron acquired in a background run by the CAST-M10 detector in the RoI, using the AFTER-based electronics. For comparison, the embedded figures show the same views for a typical $^{55}$Fe calibration event. }
\label{fig:fig2}
\end{figure}

The 3D reconstruction of an event is thus possible using the charge collected by every pad (strip) and the detector decoding. The spatial coordinates, $x$ and $y$, are determined by the integrated charge collected by each strip, while the relative $z$ position is determined from the temporal position of the strip pulses. The charge collection of each event is projected in both spatial and temporal directions. As an example, the resulting $xz$ and $yz$ projections acquired by the CAST-M10 detector and AFTER-based electronics are shown in Fig. \ref{fig:fig2} for a background event (calibration event of $^{55}$Fe shown in inset for comparison), the colour scale representing the total charge. X-rays tipically generate cloud-like energy depositions. After projecting the charge both in spatial and temporal directions, single compact groups of charge depostitions (clusters) are found in $x$, $y$ and $z$ axis. Events with only one cluster per axis are accepted. Then, using $^{55}$Fe calibrations runs, the observables that define a x-ray event are extracted (number of fired strips or temporal bins, pulse shape parameters, width in both spatial and temporal directions, charge balance between axis, etc.), and the selection criteria are defined. Finally, event selection is performed over the background events by applying these selection criteria. A more detailed description of the discrimination algorithms can be found in ~\cite{Aune:2013pna} and references therein.

This electronics readout system has been already used to characterize CAST-Micromegas spare detectors in different operation conditions~\cite{Iguaz:2011xj}: in laboratory test-benches to develop low background techniques (see Section \ref{sec:character}); and very recently, it has been installed in the three Micromegas detectors of the CAST experiment, showing good and stable performance (see Fig. \ref{fig:fig1}, right). In Table 1, a comparison of the background levels obtained with Gassiplex and AFTER electronics is established for two different setups. In the first one, AFTER electronics is used as readout system of a CAST spare detector (CAST-M10) in a setup of the University of Zaragoza. Meanwhile, the Gassiplex electronics was simulated by reducing the temporal information to the charge's integral. In the second, the comparison is established between CAST-M18 during 2012 and 2013 CAST data taking campaign, when Gassiplex and AFTER electronics were used as readout system respectively. In both cases, the improvement in background level due to the electronics upgrade has thus been quantified in $\sim$25\%.  The rejection power with AFTER-based readout electronics is higher than with former Gassiplex readout electronics as a consequence of the improvement in S/N ratio of the strip signals with respect to the mesh pulse (the only source of temporal information with Gassiplex electronics). This fact allows to efficiently extend the cluster analysis to the temporal direction. It is worth noting that the background levels obtained with CAST-M10 are systematically higher than those obtained with CAST-M18 in equivalent shielding conditions. This is due to the better performance of CAST-M18 with respect to CAST-M10, which it is a much older detector with some strips not electronically instrumented for being in shortcut with the mesh.

% In addition, the improvement in the signal to noise ratio respect to former electronics could result in an increase of the signal efficiency (i.e. the efficiency with which a x-ray is positively identified by the discrimination algorithms) below 2 keV, which is of fundamental interest for low energy rare event detection.

\begin{table}[b]
\label{tab:bkg}
\centering
$$
\begin{array}{l|ccc}
  {\mbox{Setup}}&{\mbox{\hspace{2mm} Acquisition type}}&{\hspace{2mm} \mbox{ Run time (hours)}}&{\mbox{ \hspace{2mm} Bkg. ([2-7] keV)}}\\ 
  \hline
  {\mbox{CAST-M10}}&{\mbox{Gassiplex}}&{303.1}&{2.07 \pm 0.19}\\
  {\mbox{CAST-M10}}&{\mbox{AFTER}}&{303.1}&{1.60 \pm 0.12}\\
  \hline
  {\mbox{CAST2012-M18}}&{\mbox{Gassiplex}}&{2265.5}&{1.66 \pm 0.05}\\
  {\mbox{CAST2013-M18}^{\ast}}&{\mbox{AFTER}}&{445.9}&{1.23 \pm 0.10}\\
\end{array}
$$
\caption{Comparison of the background levels obtained in the CAST RoI between Gassiplex and AFTER electronics for two different setups. Final background levels expressed in units of $10^{-6}$ \ckcs and statistical errors given as 1$\sigma$. ($^\ast$) measured with Al cathode, rest of measurements using a Cu one.} 
\end{table}

%section: VETOS Upgrade
\section{Cosmic veto implementation and results}\label{sec:veto}

\begin{figure}[htb!]
\centering
\includegraphics[height=0.35\textwidth]{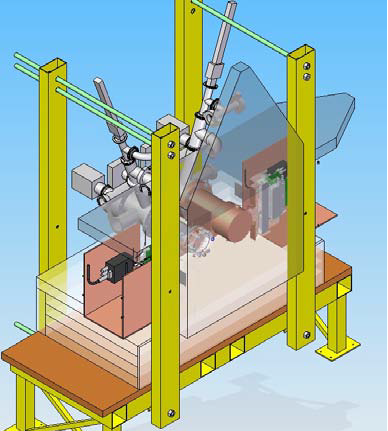} \hspace{7mm}
\includegraphics[height=0.35\textwidth]{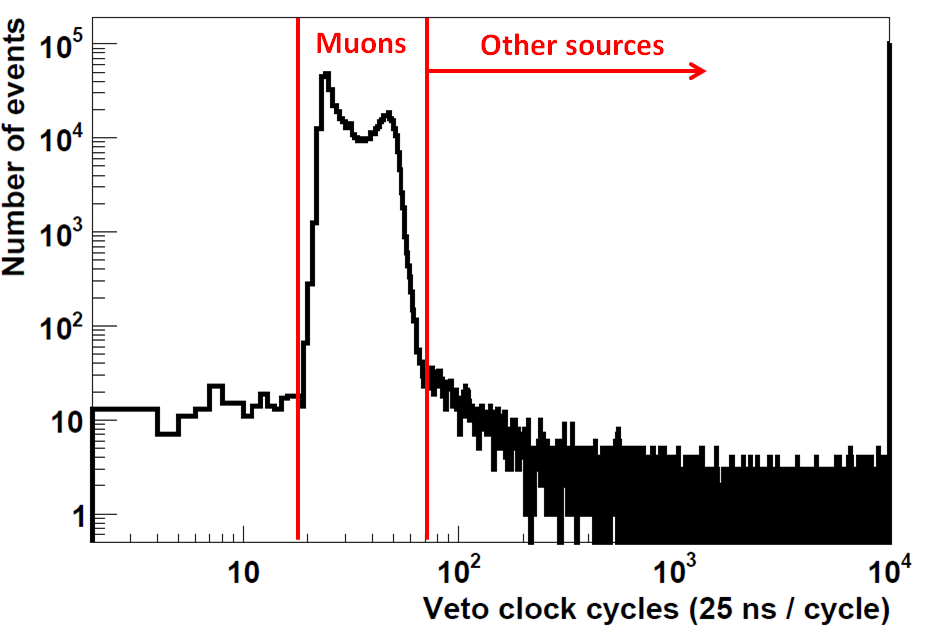}
\caption{Left: schematic view of the scintillator vetos configuration on the sunset side of the CAST magnet. The two plastic scintillator vetos on the top and back of the detectors (horizontally and vertically displayed, repectively) are arranged to cover as much solid angle as possible. Right: time difference between the signal in the muon veto and the delayed Micromegas trigger.  }
\label{fig:fig6}
\end{figure}

In the 2012 upgrade of the sunset Micromegas, a preliminary scintillator veto with ~44\% geometrical efficiency for cosmic muons was installed. It produced a reduction of $\sim$25\% in the background level. The event selection uses the recorded time difference between the signal in the muon veto and the delayed Micromegas trigger. The time difference distribution of the muon-induced events (see Fig. \ref{fig:fig6}, right) is a narrow window that represents the maximum drift time of the electrons in the chamber. Contrarily, the time difference distribution of the remaining background events follows a largely uniform distribution. The contribution of cosmic muons to background and its possible mitigation by means of higher efficient vetos has been evaluated in dedicated experimental setups.  It was found that further reductions in background level were possible. Therefore, a higher-efficiency system based on two plastic scintillators was designed, manufactured and installed in the experiment (for a schematic view of the design see Fig. \ref{fig:fig6}, left) during the 2013 summer. This installation required some changes in the allocation of the vacuum elements of the sunset line in order to increase the geometrical coverage above 90\%. The results of the tests and final setup in CAST are summarized in Table 2.

\begin{table}[ht]
\label{tab:table2}
\centering
$$
\begin{array}{l|cccc}
  {\mbox{}}&{\mbox{ \hspace{2mm} Coverage }}& {\mbox{ \hspace{2mm} Bkg. [2-7] keV }} & {\mbox{ \hspace{2mm} Bkg. [2-7] keV}}  & {\mbox{ \hspace{2mm} Reduction}}\\ 
  {\mbox{Setup}}&{\mbox{ \hspace{2mm} (\%) }}& {\mbox{ \hspace{2mm} Before veto cut }} & {\mbox{ \hspace{2mm} After veto cut}}  & {\mbox{ \hspace{2mm} (\%)}}\\ 
  \hline
  {\mbox{CAST2012-M18}^{\ast}}&{\mbox{    44   }}&{   1.66 \pm 0.05   }&{    1.29 \pm 0.05      }  &{   22 \pm 3   } \\
  {\mbox{CAST-M10}}&{\mbox{        75   }}&{   1.60 \pm 0.12   }&{    0.82 \pm 0.08      }  &{   48 \pm 5   }  \\
  {\mbox{CAST2013-M18}^{\ast\ast}}&{\mbox{    95   }}&{   1.23 \pm 0.10   }&{    0.66 \pm 0.07      }  &{   46 \pm 6   }  \\
\end{array}
$$
\caption{Background levels obtained by different detectors/setups with the same shielding. Final background levels expressed in units of $10^{-6}$ \ckcs and statistical errors given as 1$\sigma$. ($^\ast$) Gassiplex readout electronics used. Rest of measurements done with AFTER-based electronics. ($^\ast$$^\ast$) measured with Al cathode, rest of measurements using a Cu one.} 
\end{table}

\begin{figure}[htb!]
\centering
\includegraphics[height=0.28\textwidth]{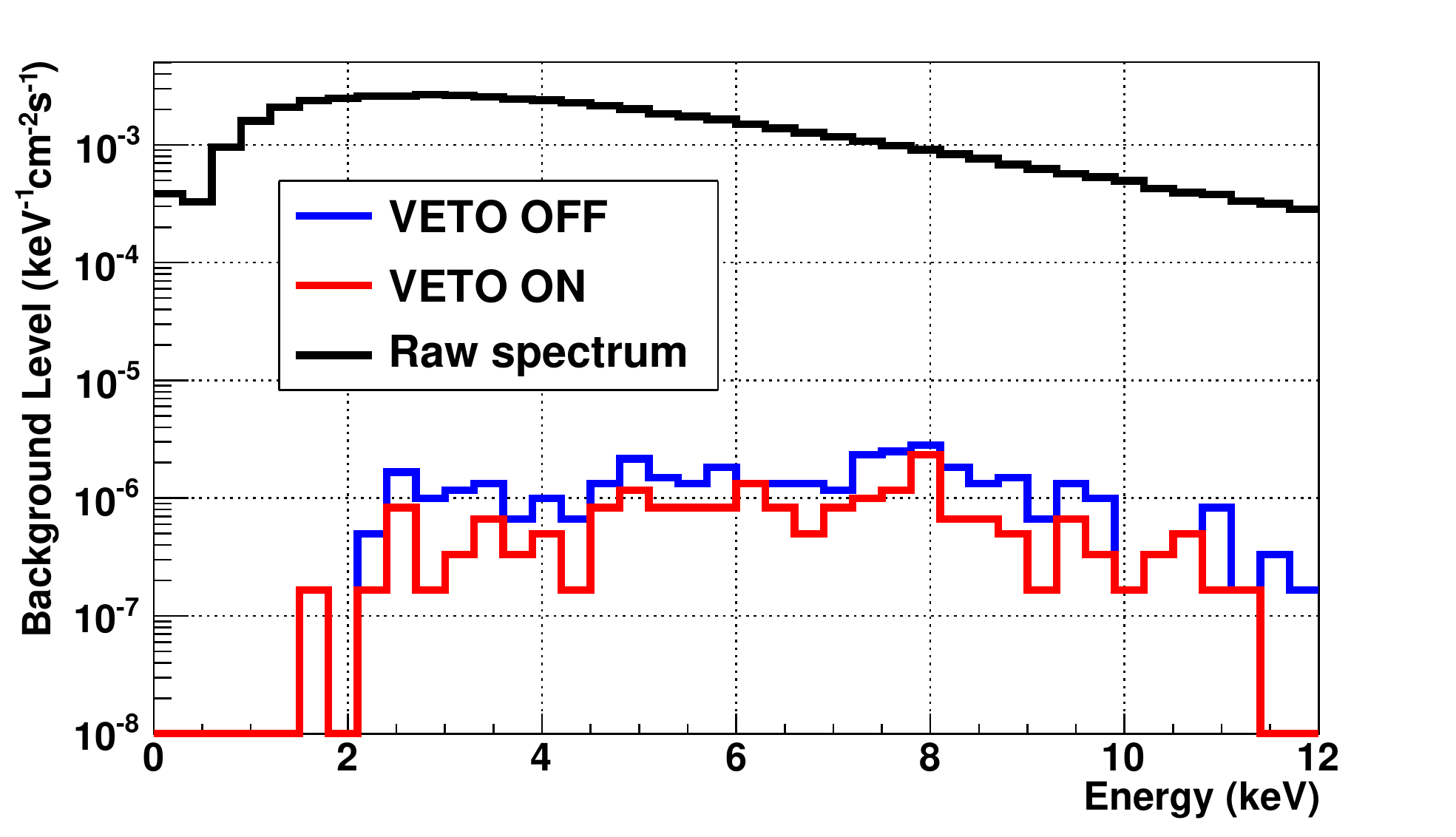} 
\includegraphics[height=0.28\textwidth]{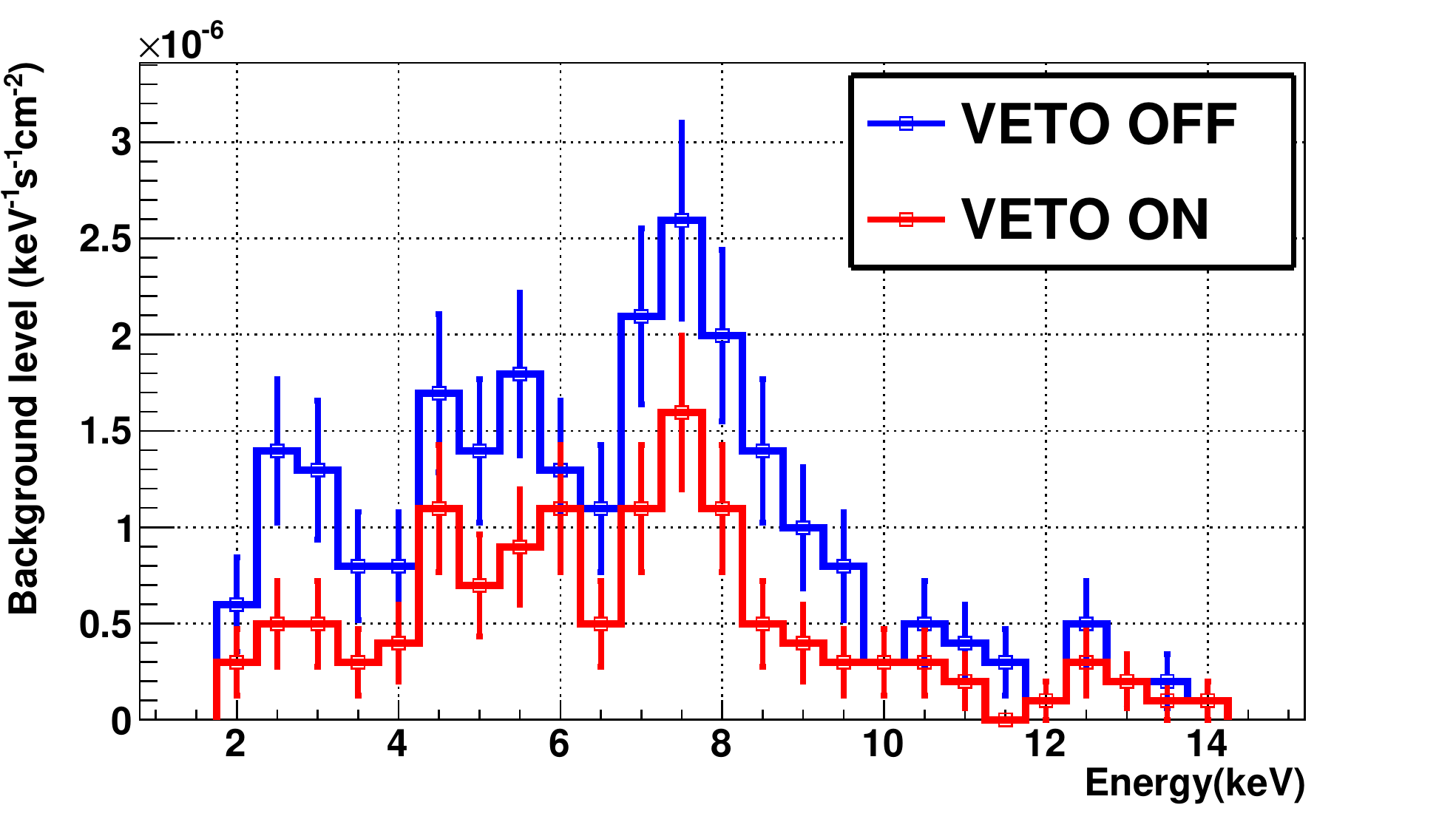} 
\caption{Background energy spectra of the CAST2013-M18 detector installed on the sunset side of the CAST magnet during the first $\sim$446 hours of the 2013 data taking campaign before (blue) and after (red) the application of the veto cut. On the left, the background level is represented in logarithmic scale to ilustrate the raw background suppression, while on the right the attention is focused on the effect of the veto cut~\cite{GCantatore:SPSC2013}.}
\label{fig:fig7}
\end{figure}

 As the Table 2 shows, the muon veto reduced the background level from (1.29 $\pm$ 0.05)$\times$10$^{-6}$  to (6.6 $\pm$ 0.7)$\times$10$^{-7}$ \ckcs in the CAST RoI ~\cite{GCantatore:SPSC2013}. On the left of Fig. \ref{fig:fig7} the raw and final background energy spectra before and after the application of the veto cut are represented for the 2013 sunset CAST-M18 data ~\cite{GCantatore:SPSC2013}. Note that the background suppression achieved by the discrimination algorithms is of around three orders of magnitude. On the right of Fig. \ref{fig:fig7}, the same final spectra are represented in linear scale, showing more clearly the effect of the veto: the suppression of the copper fluorescence events at $\sim$8 keV and the corresponding escape peak and its complementary (5 and 3 keV respectively). The background level is for the first time below $\sim$10$^{-6}$ \ckcs, a milestone in the CAST-Micromegas ultra-low background program and a step forward to the levels required by IAXO. The optimization of the discrimination criteria for the AFTER-based electronics is under development, so even lower values may be achievable. The reduction in background level represents an improvement of a factor 2 with respect to 2012 sunset Micromegas. The causes of the improvement are shared between the upgrade in the electronics described in Section \ref{sec:electronics} and the cosmic veto upgrade.

%a preliminary analysis of the first $\sim$446 hours of the 2013 sunset CAST-M18 data with this setup shows that, keeping the same signal efficiency as for the 2012 data,the background level after the application of the discrimination algorithms has been reduced from

It is worth noting that this level has been obtained with TPC-cathodes made out of aluminum instead of the more radiopure copper, due to technical reasons. The contribution to background of the aluminum cathode was evaluated in a dedicated setup in the Canfranc Underground Laboratory (LSC), and was assessed to be at a level $\sim$3--5$\times$10$^{-7}$ \ckcs~\cite{Aune:2013pna}. The replacement of this cathode by a copper one will very likely imply an additional improvement that will allow approaching the current best background level achieved by a Micromegas detector underground ($\sim$10$^{-7}$ \ckcs ~\cite{Aune:2013pna}). With this replacement, the reduction in background level due to the veto of the sunset setup will very likely be higher than 70\%, instead of the 46\% currently obtained.

%section: CHAMBER Upgrade
\section{Design and $^{55}$Fe characterization of new CAST-Micromegas detectors} \label{sec:character}

\subsection{Design and implementation}
\begin{figure}[htb!]
\centering
\includegraphics[height=0.28\textwidth]{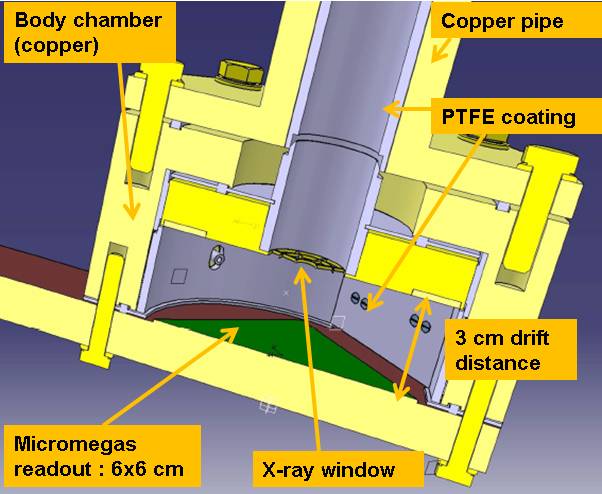} \hspace{6mm}
\includegraphics[height=0.28\textwidth]{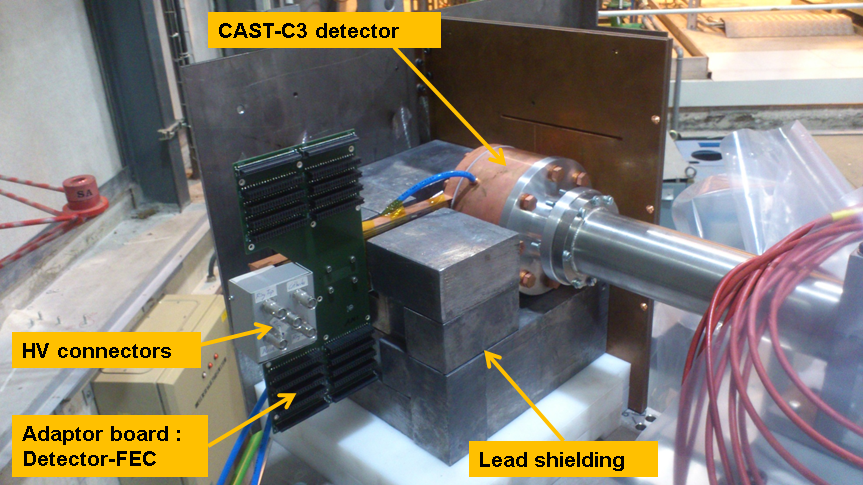}
\caption{Left: schematic view of the new detector design. Right: new detector installed in the vacuum line of the sunrise side of CAST magnet, partially shielded with lead.}
\label{fig:fig3}
\end{figure}

A newly designed CAST-Micromegas detector has been installed at the sunrise side in 2013. The improvements in the design are based on the current picture of the background model~\cite{Aune:2013pna}. Therefore, the goal of the upgrade is to produce a detector with the highest possible radiopurity and shielding, and with improved TPC properties. All the detector materials have been screened and selected for radiopurity: pure copper is the base material of the chamber structure, polytetrafluoroethylene (PTFE) is used for electric insulation and sealing, and polyimide is used as substrate for printed circuits. All these materials have well-known low intrinsic radioactivity ~\cite{Susana:mpgd2013}, and have replaced the less radiopure components of previous designs. The gas chamber, Faraday cage and inner shielding are completely integrated in the new detector design (see Fig. \ref{fig:fig3}, left). The former plexiglas body of the TPC has been replaced by a 18 mm thick copper one. This fact allows to relax some mechanical constraints enabling a thicker and more efficient lead shielding. The support base on which the Micromegas is glued is also made of copper, minimizing the signal extraction outlet. It is worth noting that the printed board brings out of the shielding not only the Micromegas signals (mesh and strips) but also the drift high voltage connections. The electrical connections, being a potential source of radioactivity ~\cite{Susana:mpgd2013}, are in this way moved away from the detector body, without the need for exit points in the shielding. A new cylindrical drift field shaper has alplso been designed, printed on a flexible multilayer circuit with polyimide as substrate. The outer side of the circuit is used to bring the HV connections from the detector board to the field-shaper traces and to the drift cathode. An inner  PTFE foil blocks the fluorescence from these traces without disturbing the field lines. 

The readout pattern has been barely modified with respect to previous CAST-Micromegas designs. The new detectors still use the microbulk technology. The properties and performance of microbulk Micromegas demonstrate that they are the optimum choice for CAST purposes~\cite{Dafni:2012fi,Abbon:2007ug}. The amplification gap is kept at 50 $\mu$m, while the number of strips per axis is slightly increased (from 106 to 120) in a 60$\times$60 mm$^{2}$ readout area (500 $\mu$m pitch). The main novelty is that the printed board contains not only the 120$\times$120 anode channels but also the traces for the mesh and the drift high voltages. The anode traces end in a multi-pin connector situated $\sim$10 cm away from the detector chamber. An adaptor board and high-density flat flexible cables are used to bring the signals from the detector to the FEC, which can now be situated tens of centimeters far from the detector chamber (see Fig. \ref{fig:fig3}, right). This fact minimizes the potential contribution to the background level due to the radioactivity of electronic components given that rigid PCBs main insulating component (FR4) is know to be non radiopure ~\cite{Susana:mpgd2013}. A dedicated setup in the LSC facilities is being installed to quantify the contribution of the electronic components to the background level.

In the future, towards 2014, the installation of a dedicated x-ray focusing device~\cite{ThPapaevangelou:SPSC2012} coupled to the Micromegas detector will enable the use of a much smaller window, given that x-rays produced by back-conversion of solar axions in the CAST magnet will be focused on a spot of a few mm$^{2}$, to be compared with the current sensitive area of 14.5 cm$^{2}$. Therefore, the outer diameter of the vacuum pipe that couples the detector to the x-ray focusing device will be also much narrower. This fact, together with the shift of the focal point with respect to the magnet axis introduced by the optics, removes many of the mechanical constraints limiting the efficiency of the shielding. The lead shielding of the sunrise CAST-Micromegas detector for 2014 will be 10 cm thick, instead of the $\sim$5 cm of the present setup, and it will be extended a few tens of centimeters along the vacuum pipe providing better solid angle coverage. In addition, the vacuum pipe will be made out of copper (with a PTFE foil in the inner surface to block the 8 keV fluorescence) instead of stainless steel (whose fluorescence lines are in the 5-7 keV range, inside the CAST RoI).

\subsection{Characterization with a $^{55}$Fe source}
\begin{figure}[htb!]
\centering
\includegraphics[height=0.32\textwidth]{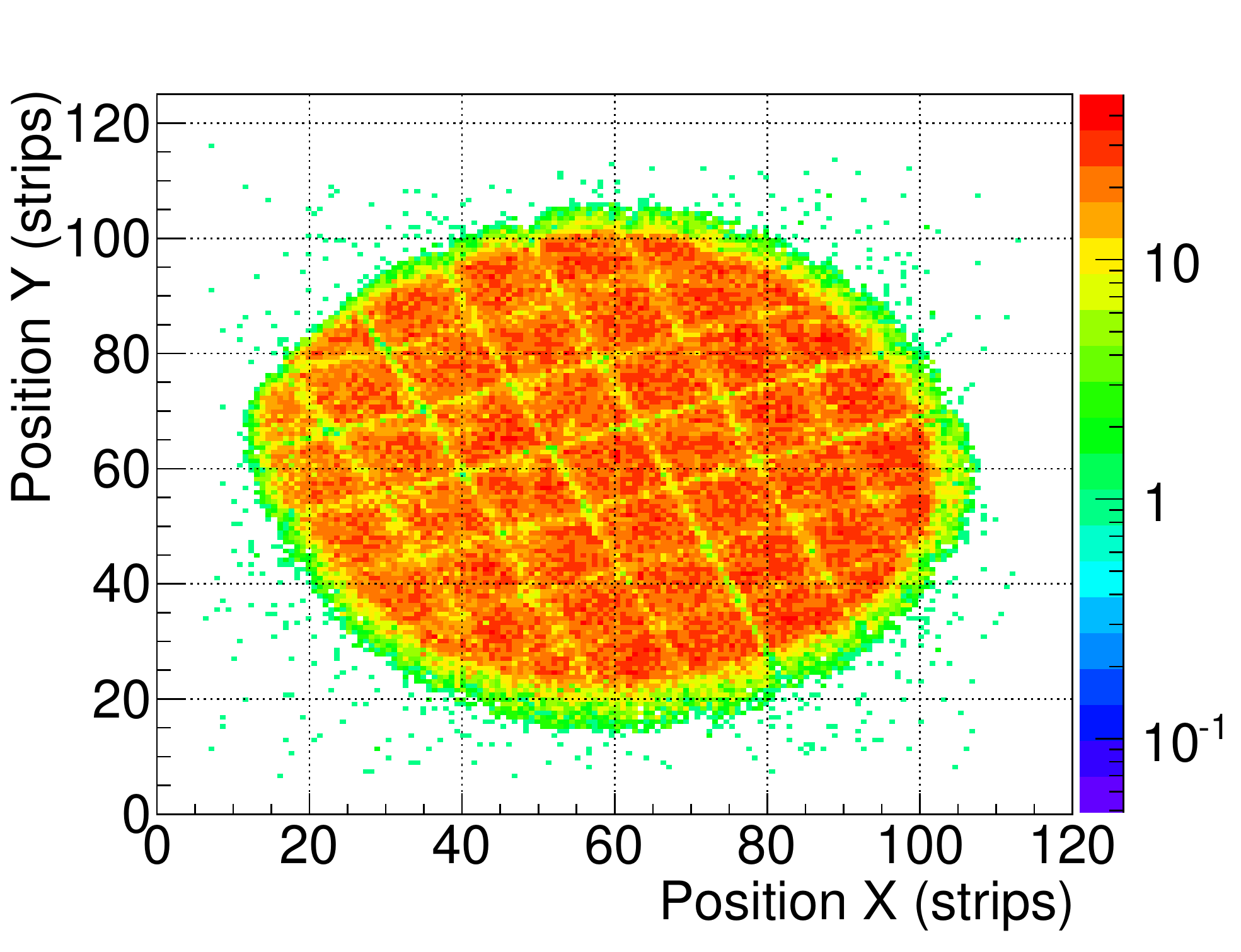} 
\includegraphics[height=0.30\textwidth]{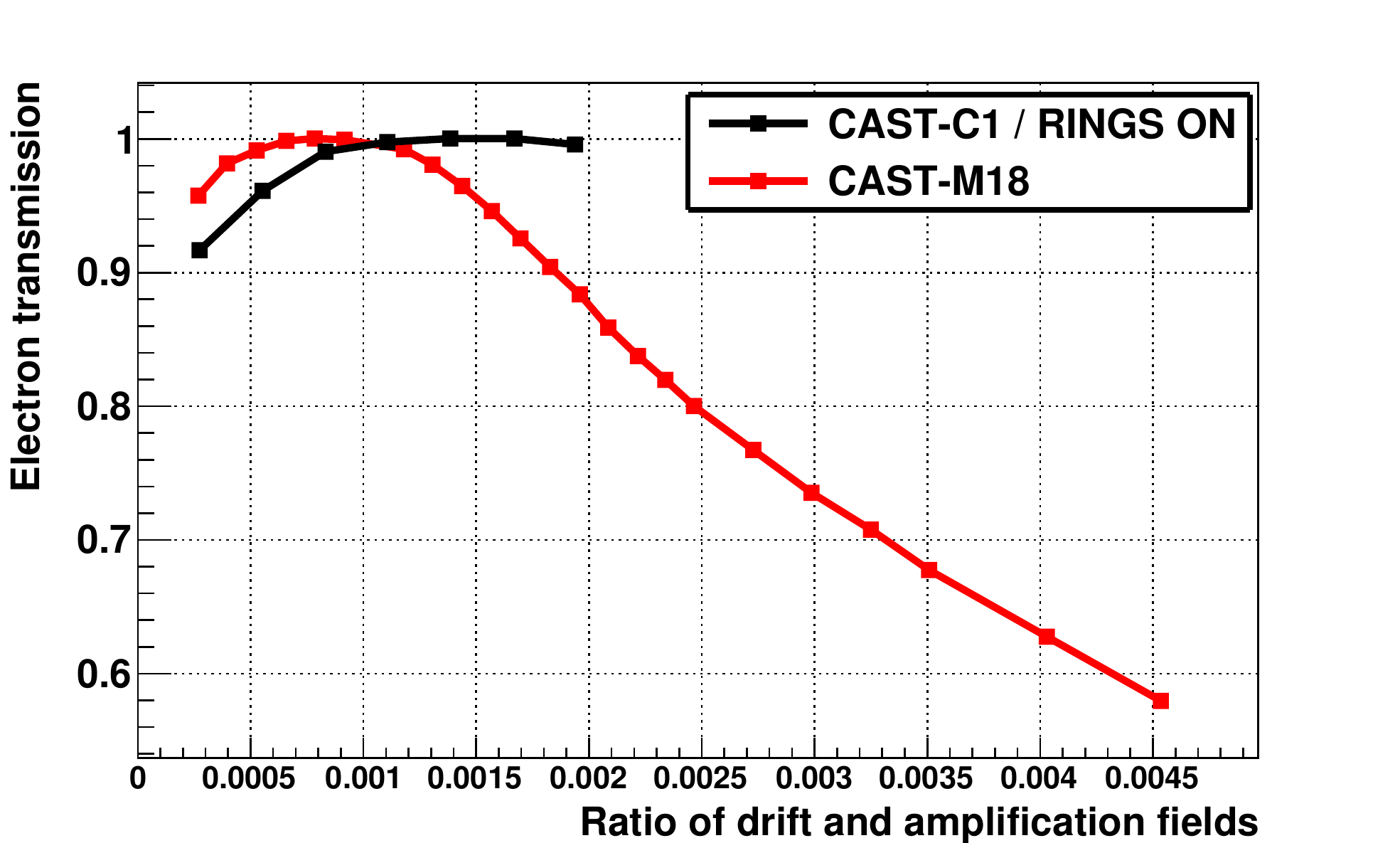}
\caption{Left: hitmap distribution generated by $^{55}$Fe calibration events of the CAST-C3 detector (strip pitch is 500 $u$m). The orthogonal shadowed lines correspond to strong-back support of the x-ray entrance window. Right: dependence of the electron transmission with the ratio of drift and amplification fields for the new microbulk CAST-C1 detector at 1 bar. A curve of the old-type CAST-M18 detector is also shown for comparison. The gain has been normalized to the maximum of each series.}
\label{fig:fig4}
\end{figure}

Three of these new detectors (CAST-C1, -C2 and -C3) have been built and characterized using the 5.9 keV x-rays of a $^{55}$Fe source in Ar+2.3\%iC$_{4}$H$_{10}$ at 1 bar. All the detectors have a sensitive surface large enough to cover the axion-sensitive area, i.e. the magnet cold bore, 14.5 cm$^{2}$ (as an example see CAST-C3 hitmap on the left of Fig. \ref{fig:fig4}). Firstly, the drift field was varied at a fixed mesh voltage (300 V).  The voltages applied to the cathode typically
ranged from 350 to 650 V and the bias of the field shaper was varied accordingly to maintain a uniform electric field. These tests were done to obtain which is the optimum ratio of drift-to-amplification field for a maximum mesh transparency to primary electrons. The electron transmission curve obtained for one of these detectors is shown on the right of Fig. \ref{fig:fig4}, compared to the one obtained for the CAST-M18 detector. Although the range of drift fields over which we conducted the tests is reduced due to technical difficulties when applying high voltages, a clear plateau of maximum electron transmission is observed for all the detectors. At high drift fields, the mesh stops being transparent for primary electrons in both detectors, but the plateau is larger in the new one. This is fact may be attributed to the fact that the spacing between readout pads has been notably reduced in the new detectors (from 50 to 30 $u$m).

\begin{figure}[htb!]
\centering
\includegraphics[height=0.28\textwidth]{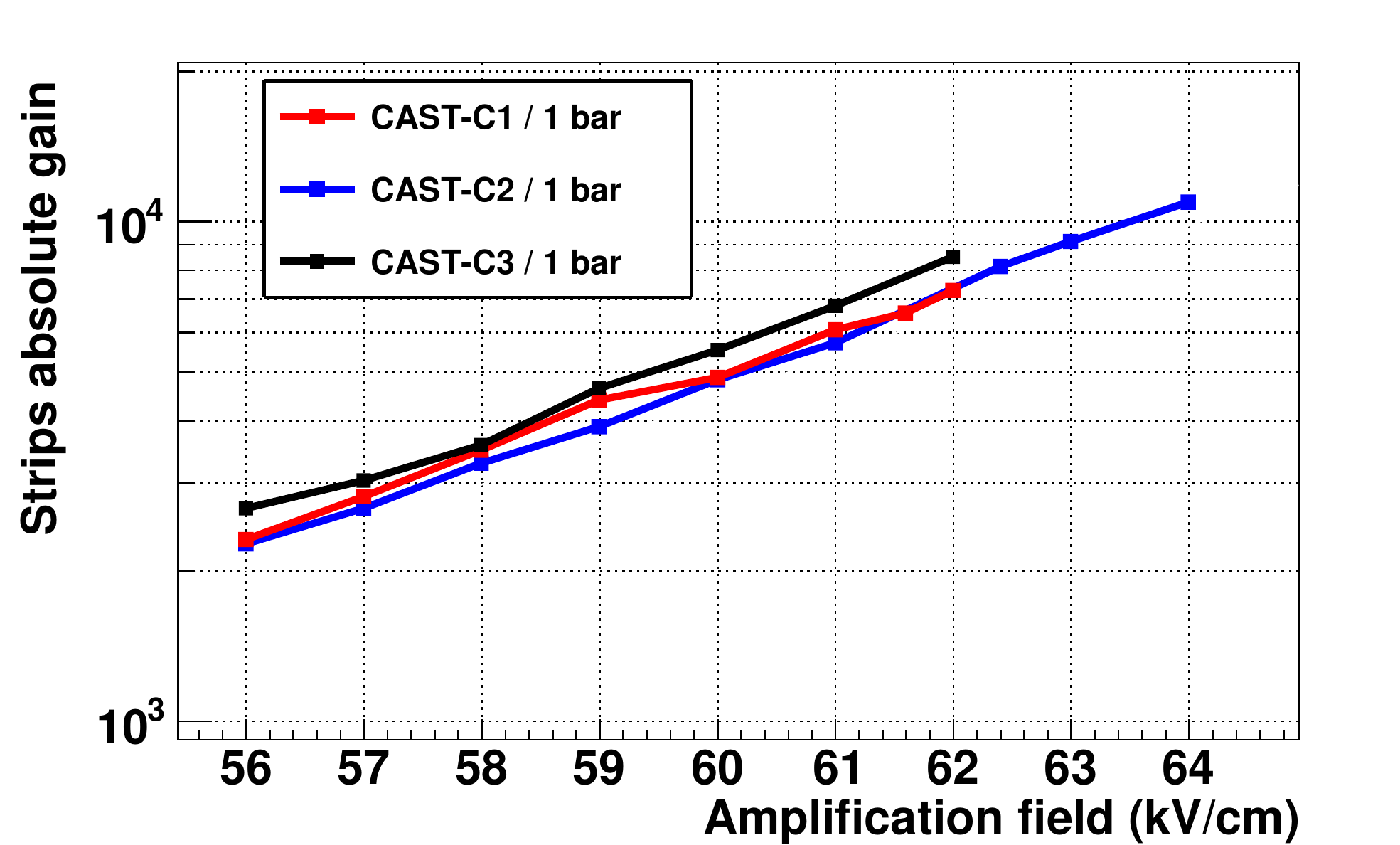} 
\includegraphics[height=0.28\textwidth]{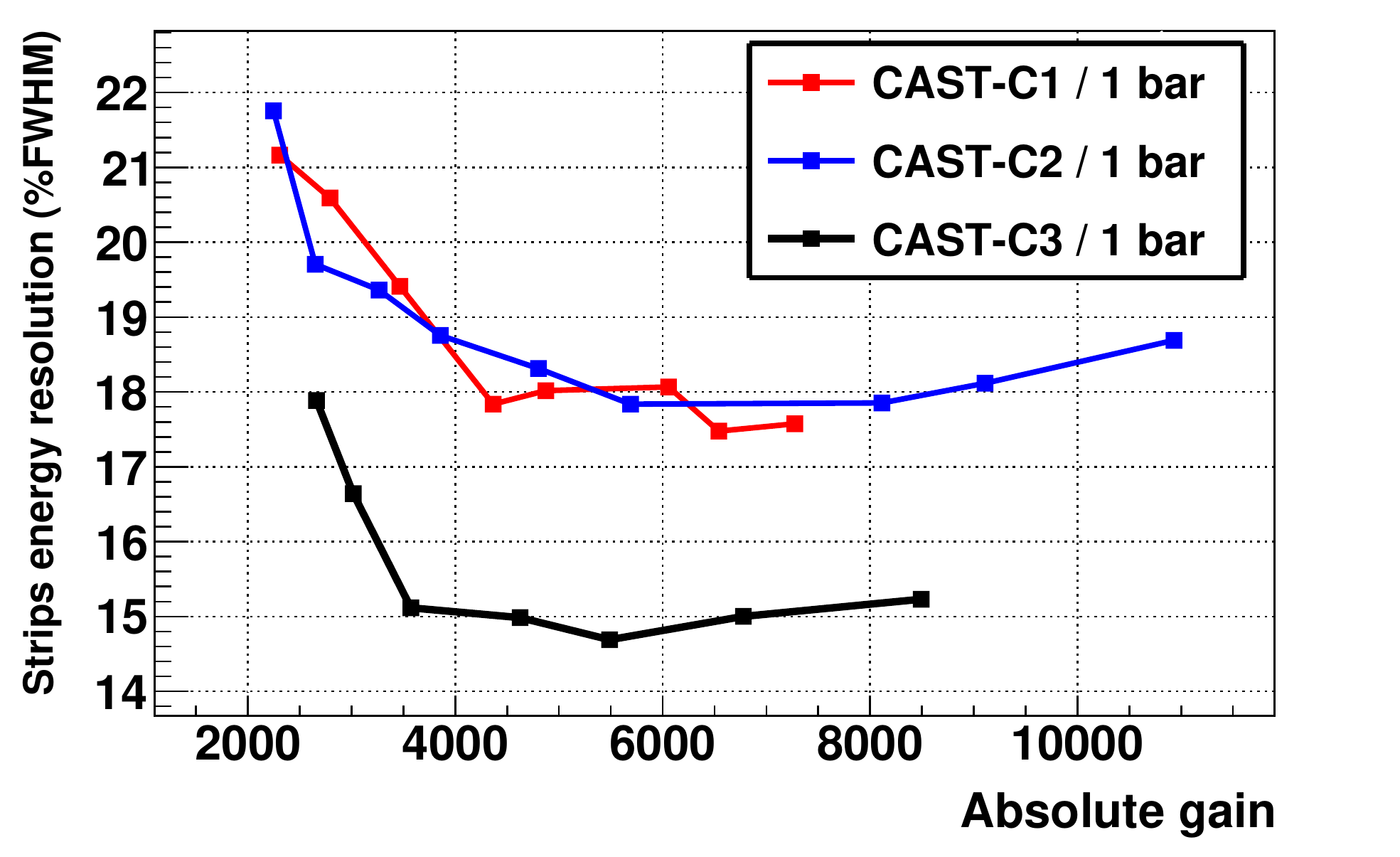}
\caption{Left: dependence of the absolute gain with the amplification field for the CAST-C1, -C2 and -C3 detectors in Ar+2.3\%iC$_{4}$H$_{10}$ at 1 bar. Right: dependence of the strips energy resolution with the absolute gain. }
\label{fig:fig5}
\end{figure}

Once the optimum ratio of drift-to-amplification field is determined for every detector, the mesh voltage is varied from $\sim$280 V to $\sim$310 V to obtain the gain curves shown in the left of Fig. \ref{fig:fig5}. The curves reflect the dependence of the amount of charge collected by the strips for a 5.9 keV event with the mesh voltage. The maximum gain before the spark limit is around 10$^{4}$. The dependence of the strips energy resolution with the amplification field, and more specifically with the gain, is shown on the right of Fig. \ref{fig:fig5}. At low gain, the energy resolution worsens due to a worse S/N ratio; while at high gains, the energy resolution degrades due to larger gain fluctuations~\cite{Iguaz:2012ur}.The CAST-C3 detector shows an energy resolution considerably better than the rest of the detectors of the series, reaching 14.7\% FWHM at 5.9 keV. This is due to the fact that CAST-C1 and CAST-C2 detectors present several strips shortcircuited among them, due to a failure in the manufacturing process. Even if the best values observed for microbulk detectors are $\sim$11\% FWHM\footnote{Note that this value was obtained in Ar+5\% C$_{4}$H$_{10}$, a more favorable mixture for 50 $\mu$m gap microbulk Micromegas}~\cite{Cebrian:2010nw,Iguaz:2011xi}, the CAST-C3 energy resolution is a very remarkable value given that a large area is illuminated and no fiducialization has been applied, showing the high uniformity of the charge amplification over the detector surface. 

%section:  X-ray beam PIXE
\section{Micromegas characterization in a x-ray beam line} \label{sec:xraybeam}

\begin{figure}[htb!]
\centering
\includegraphics[height=0.28\textwidth]{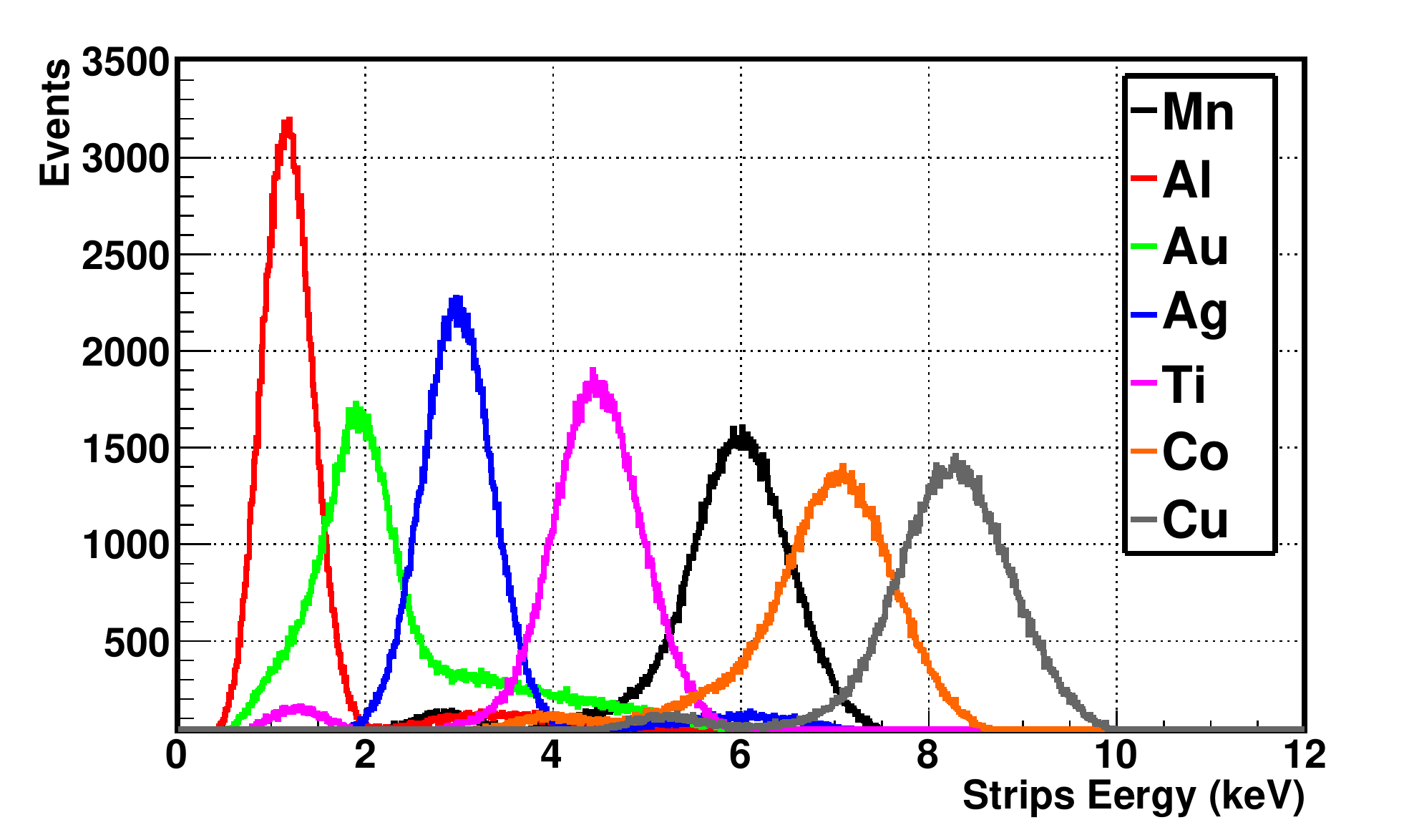} 
\includegraphics[height=0.28\textwidth]{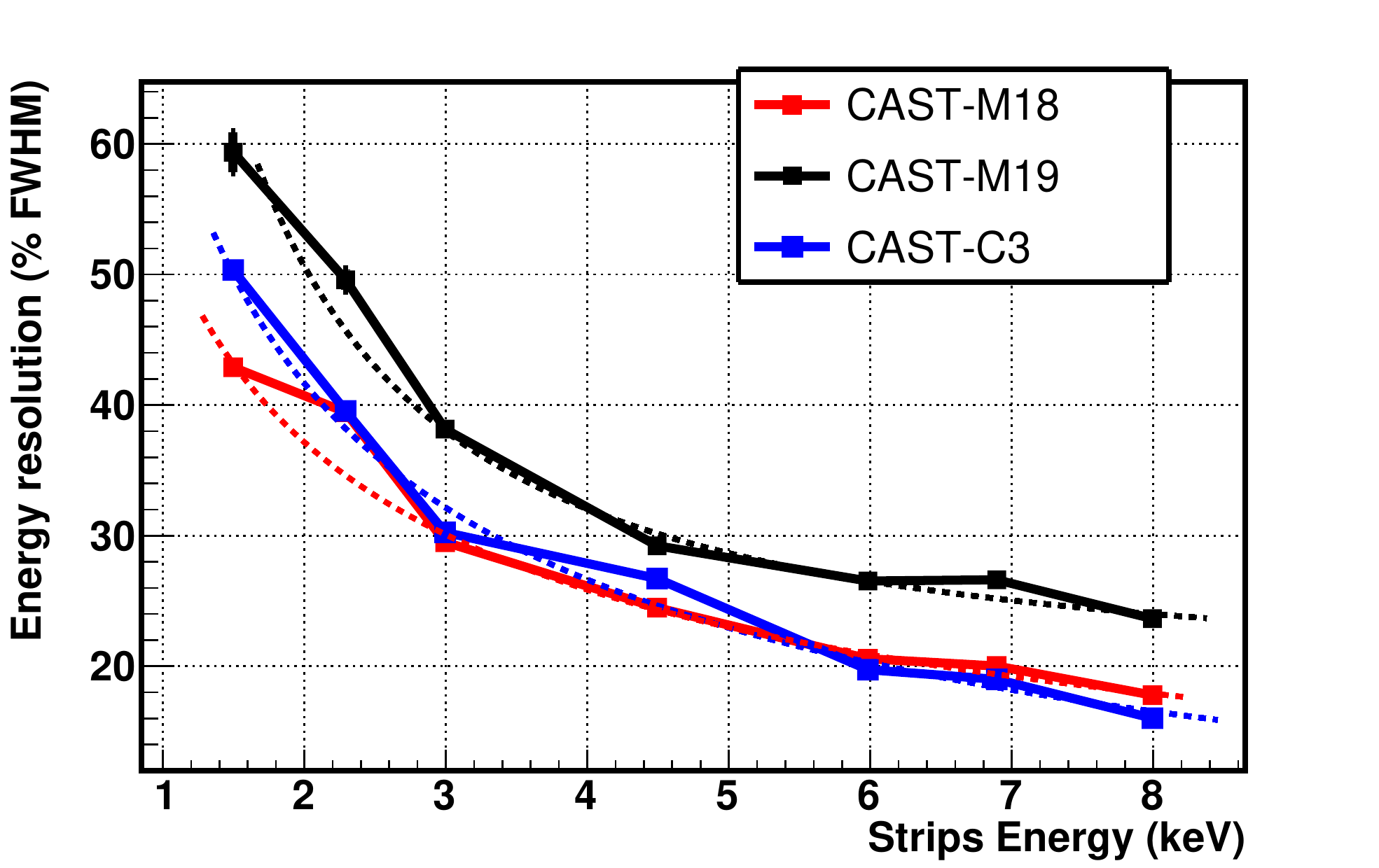} 
\caption{Left: calibration energy spectra of the  CAST-M18 detector installed in the x-ray beam of the CAST Detector Laboratory. Several materials were used as targets of the PIXE system to better scan the CAST energy RoI. Right: dependence of the energy resolution registered by the strips with the incident x-ray energy for three CAST-Micromegas detectors.}
\label{fig:fig8}
\end{figure}

A set of dedicated measurements using an electron beam based on PIXE (Particle Induced X-ray Emission) in the CAST Detector Laboratory at CERN~\cite{theotesis} has allowed to calibrate the CAST-Micromegas detectors at several x-ray energies, using the fluorescence lines of different target materials  (see Fig. \ref{fig:fig8}, left). These measurements provide a better characterization and understanding of the detector response as a function of the incident x-ray energy. The dependence of the energy resolution of the strips signals with the incident x-ray energy is shown on the right of Fig. \ref{fig:fig8} for three CAST-Micromegas detectors. The experimental points show good agreement with the scaling given by the parametrization: $\frac{\sigma}{E} = \frac{a}{\sqrt{E}} + \frac{b}{E}$, where the first term represents the statistical fluctuation of the electrons generating the signal and the second is the calibration term.  

\begin{figure}[htb!]
\centering
\includegraphics[height=0.28\textwidth]{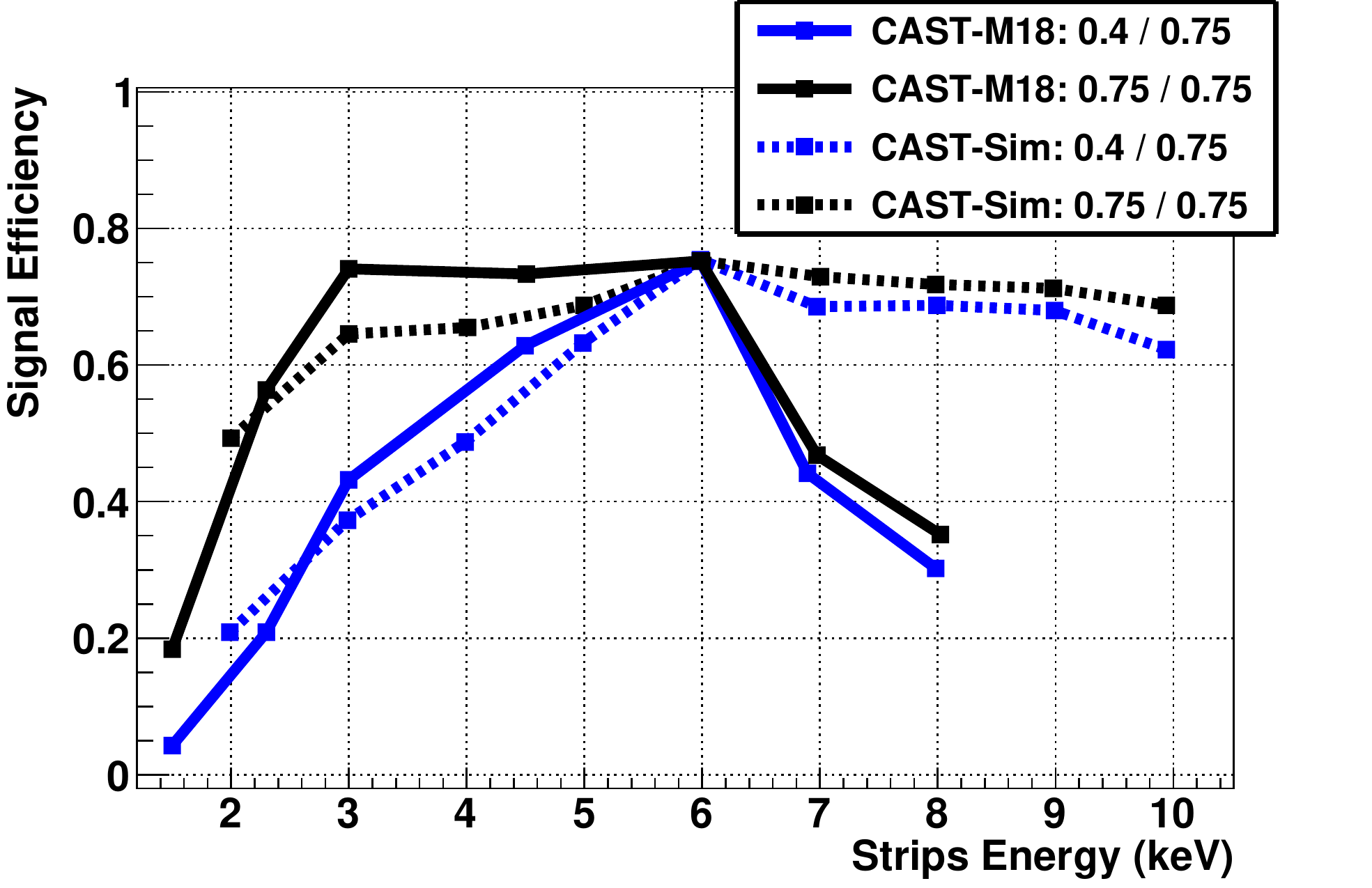} 
\includegraphics[height=0.28\textwidth]{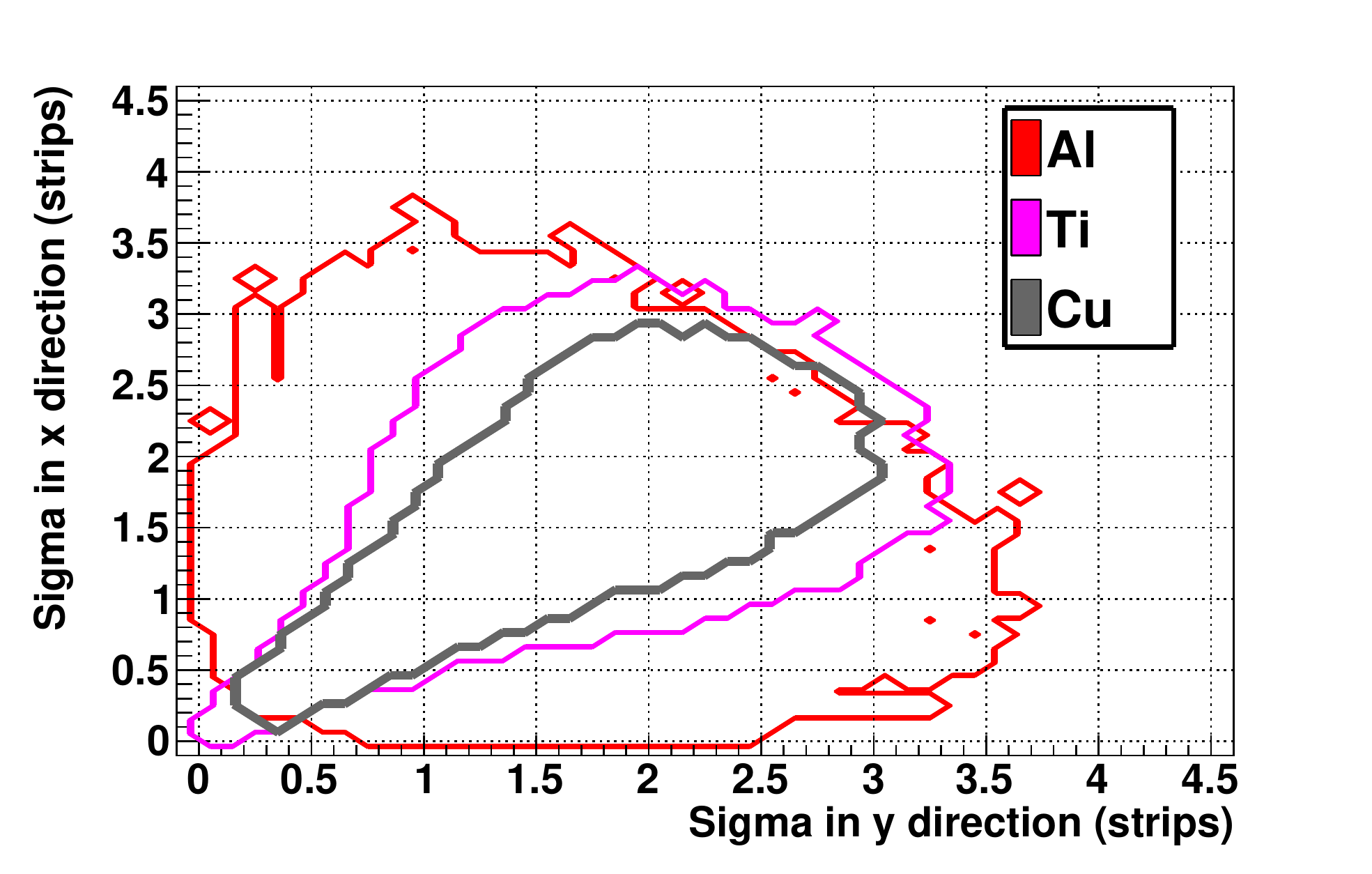} 
\caption{Left: comparison between the simulated and experimentally measured signal efficiency for two different sets of fixed efficiencies at 6 and 3 keV, the main and escape peak of a $^{55}$Fe calibration source. Right: distribution of  usual observables used in the CAST analysis for three different target materials. }
\label{fig:fig9}
\end{figure}

The applications of these type of measurements are multiple. On one hand, the integrated signal efficiency over the CAST RoI, a parameter that enters into the calculation of the upper limit to the axion-to-photon coupling constant, can be evaluated on basis of experimental data. Formerly, this parameter was calculated using MonteCarlo simulations of the x-ray energy deposition and electronic signal generation on the electrodes of the Micromegas detector~\cite{alfredotesis}.  A comparison between the simulated and the experimentally measured signal efficiency  is shown on the left of Fig. \ref{fig:fig9}, using an analysis that  guarantees fixed efficiencies for the two energies available with the typical $^{55}$Fe calibrations (3 and 6 keV), which are daily done during the running periods of the experiment. The results, although dependent on the discrimination algorithms applied, generically show that simulated and experimental data reasonably agree below 6 keV, but they highly differ above that energy. While the signal efficiency for the simulated data decrease very slightly above 6 keV, the experimental efficiency drops much faster. The divergence is attributed to an over idealistic signal generation in the simulated data. On the other hand, it allows to study the dependence of the distribution of the observables used in the CAST analysis with the incident x-ray energy (an example is shown on the right of Fig. \ref{fig:fig9}). These dependences may be applied to the development of more efficient energy-dependent discrimination algorithms.

% section: PROSPECTS
\section{Future prospects}\label{sec:proscpects}

The vacuum run in CAST in 2014 will allow testing a dedicated x-ray focusing device coupled to a Micromegas detector. This  means merging the two main CAST innovations regarding low energy x-ray detection, and will yield an increase of sensitivity. The x-rays coming from the CAST magnet will be focused in a spot of a few mm$^{2}$, in comparison with the current sensitive area of 14.5 cm$^{2}$.

\begin{figure}[htb!]
\centering
\includegraphics[height=0.6\textwidth]{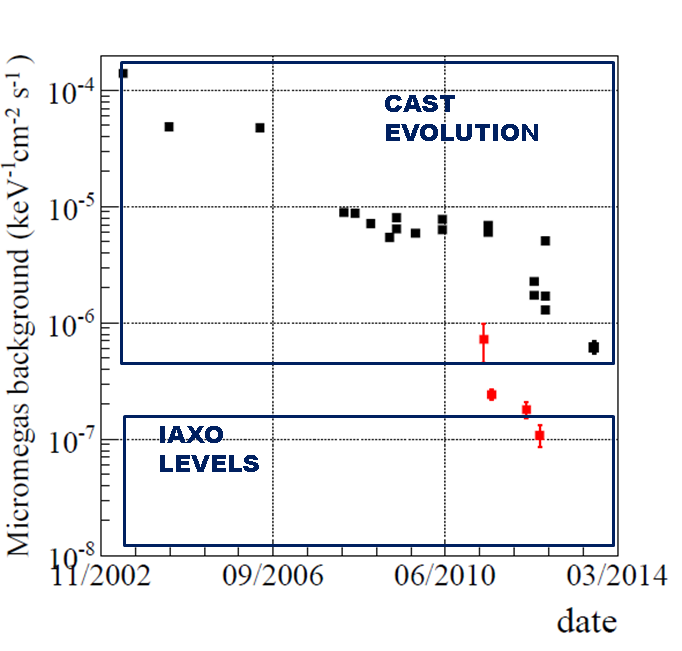} 
\caption{Black squared points represent background levels of the Micromegas detectors achieved in the CAST experiment. The lower point of this series corresponds to the 2013 data taking campaign. The red points represent the background levels achieved deep underground in the LSC facilities along the years.}
\label{fig:fig10}
\end{figure}

As mentioned in Section \ref{sec:veto}, the replacement of the aluminum cathodes by the more radiopure copper ones in the sunset detectors will lead to background levels close to the best values obtained underground and to the IAXO requirements (see Fig. \ref{fig:fig10}). In order to reach these levels there are still open research lines that are being actively studied both at sea level and in deep underground setups. The underground setup at LSC has proven background levels down to $\sim10^{-7}$ \ckcs even with Gassiplex readout electronics. A future upgrade to AFTER-based electronics will allow to apply the new discrimination criteria using the extra gathered information. Besides, this upgrade will enable moving away the electronic system (not screened for radiopurity) from the detector chamber. A thicker external shielding will possibly be necessary in order to reveal a new background limit. On top of this, a new neutron-shielding based on a low $Z$ material to moderate them (typically water or polyethylene) and a neutron absorber (typically cadmium) could be tested, as well as an active muon veto system.
% In surface tests, more efficient cosmic vetos can be tested in order to tag muon-induced events by secondary particle production. 

In parallel, the development of more radiopure materials for the Micromegas manufacturing and the production of segmented mesh detectors ~\cite{Thomas_SegmentedMesh} are under study. More radiopure raw materials and mass minimization can lower the limit imposed by the intrinsic radioactivity of current detectors.

% section: CONCLUSIONS
\section{Conclusions}

The Micromegas detectors have demostrated to be an optimal technological choice for low background x-ray detection, as the evolution of CAST-Micromegas background level shows. In this document, the latest upgrades and tests towards background minimization and better understanding of the detector response have been described. Background levels below 10$^{-6}$ \ckcs have been achieved for the first time at sea level by means of a highly efficient cosmic muon veto system. The optimization of the discrimination criteria with the AFTER-based electronics is under development, so even lower values are anticipated. Nevertheless, there are still open lines for improvement. The characterization and properties of the new detectors, whose design is presented in Section \ref{sec:character}, make them the baseline design for the future low energy x-ray Micromegas detectors. The demonstrative plans towards the required levels of IAXO have been presented in Section \ref{sec:proscpects}. These plans comprise new x-ray focusing devices coupled to the Micromegas detectors, upgrades in passive and active shielding in underground and surface tests, and the use of even higher radiopure materials.

%\acknowledgments
\section*{Acknowledgements}
We want to thank our colleagues of CAST for many years of collaborative work in the experiment, and many helpful discussions and encouragement. We thank R. de Oliveira and his team at CERN for the manufacturing of the microbulk readouts. We also thank the LSC staff for their help in the support of the Micromegas setup at the LSC. Authors would like to acknowledge the use of Servicio General de Apoyo a la Investigaci\'on-SAI, Universidad de Zaragoza. F.I. acknowledges the support of the Eurotalents program. We acknowledge support from the European Commission under the European Research Council T-REX Starting Grant ref. ERC-2009-StG-240054 of the IDEAS program of the 7th EU Framework Program. We also acknowledge support from the Spanish Ministry of Economy and Competitiveness (MINECO) under contracts ref. FPA2008-03456 and FPA2011-24058, as well as under the CPAN project ref. CSD2007-00042 from the Consolider-Ingenio 2010 program. Part of these grants are funded by the European Regional Development Fund (ERDF/FEDER).

\bibliographystyle{JHEP}
\bibliography{igorbib}

\providecommand{\href}[2]{#2}\begingroup\raggedright\begin{thebibliography}{10}

\bibitem{Ringwald:2012cu}
A.~Ringwald, {\it {Searching for axions and ALPs from string theory}},
  \href{http://xxx.lanl.gov/abs/1209.2299}{{\tt arXiv:1209.2299}}.

\bibitem{Peccei:1977hh}
R.~D. Peccei and H.~R. Quinn, {\it {CP conservation in the Presence of
  Instantons}},  {\em Phys. Rev. Lett.} {\bf 38} (1977) 1440--1443.

\bibitem{Abbott:1982af}
L.~Abbott and P.~Sikivie, {\it {A Cosmological Bound on the Invisible Axion}},
  {\em Phys.Lett.} {\bf B120} (1983) 133--136.

\bibitem{Dine:1982ah}
M.~Dine and W.~Fischler, {\it {The Not So Harmless Axion}},  {\em Phys.Lett.}
  {\bf B120} (1983) 137--141.

\bibitem{Preskill:1982cy}
J.~Preskill, M.~B. Wise, and F.~Wilczek, {\it {Cosmology of the Invisible
  Axion}},  {\em Phys.Lett.} {\bf B120} (1983) 127--132.

\bibitem{Zioutas:2004hi}
{\bf CAST} Collaboration, K.~Zioutas et~al., {\it {First results from the CERN
  Axion Solar Telescope (CAST)}},  {\em Phys. Rev. Lett.} {\bf 94} (2005)
  121301, [\href{http://xxx.lanl.gov/abs/hep-ex/0411033}{{\tt
  hep-ex/0411033}}].

\bibitem{Giomataris:1995fq}
Y.~Giomataris, P.~Rebourgeard, J.~P. Robert, and G.~Charpak, {\it {MICROMEGAS:
  A high-granularity position-sensitive gaseous detector for high particle-flux
  environments}},  {\em Nucl. Instrum. Meth.} {\bf A376} (1996) 29--35.

\bibitem{Andriamonje:2010zz}
S.~Andriamonje, D.~Attie, E.~Berthoumieux, M.~Calviani, P.~Colas, et~al., {\it
  {Development and performance of Microbulk Micromegas detectors}},  {\em
  JINST} {\bf 5} (2010) P02001.

\bibitem{Iguaz:2011xi}
F.~Iguaz, S.~Andriamonje, F.~Belloni, E.~Berthoumieux, M.~Calviani, et~al.,
  {\it {New developments in Micromegas Microbulk detectors}},  {\em
  Phys.Procedia} {\bf 37} (2012) 448--455,
  [\href{http://xxx.lanl.gov/abs/1110.2641}{{\tt arXiv:1110.2641}}].

\bibitem{kuster2007}
M.~Kuster, H.~Braeuninger, S.~Cebri{\'a}n, M.~Davenport, C.~Eleftheriadis,
  J.~Englhauser, H.~Fischer, J.~Franz, P.~Friedrich, R.~Hartmann, F.~H.
  Heinsius, D.~H.~H. Hoffmann, G.~Hoffmeister, J.~N. Joux, D.~Kang,
  K.~Koenigsmann, R.~Kotthaus, T.~Papaevangelou, C.~Lasseur, A.~Lippitsch,
  G.~Lutz, J.~Morales, A.~Rodriguez, L.~Strueder, J.~Vogel, and K.~Zioutas,
  {\it {The x-ray telescope of CAST}},  {\em {NEW JOURNAL OF PHYSICS}} {\bf
  {9}} ({JUN 22}, {2007}).

\bibitem{Irastorza:2011gs}
I.~G. Irastorza, F.~Avignone, S.~Caspi, J.~Carmona, T.~Dafni, et~al., {\it
  {Towards a new generation axion helioscope}},  {\em JCAP} {\bf 1106} (2011)
  013, [\href{http://xxx.lanl.gov/abs/1103.5334}{{\tt arXiv:1103.5334}}].

\bibitem{Irastorza:1567109}
I.~G. Irastorza, {\it The international axion observatory iaxo. letter of
  intent to the cern sps committee.},  Tech. Rep. CERN-SPSC-2013-022.
  SPSC-I-242, CERN, Geneva, Aug, 2013.

\bibitem{Aune:2013pna}
S.~Aune, J.~Castel, T.~Dafni, M.~Davenport, G.~Fanourakis, et~al., {\it {Low
  background x-ray detection with Micromegas for axion research}},
  \href{http://xxx.lanl.gov/abs/1310.3391}{{\tt arXiv:1310.3391}}.

\bibitem{Santiard:1994ps}
J.~Santiard, W.~Beusch, S.~Buytaert, C.~Enz, E.~Heijne, et~al., {\it {Gasplex:
  A Low noise analog signal processor for readout of gaseous detectors}}, .

\bibitem{Baron:2008zza}
P.~Baron, D.~Calvet, E.~Delagnes, X.~de~la Broise, A.~Delbart, et~al., {\it
  {AFTER, an ASIC for the readout of the large T2K time projection chambers}},
  {\em IEEE Trans.Nucl.Sci.} {\bf 55} (2008) 1744--1752.

\bibitem{Baron:2010zz}
P.~Baron, D.~Besin, D.~Calvet, C.~Coquelet, X.~De~La~Broise, et~al., {\it
  {Architecture and implementation of the front-end electronics of the time
  projection chambers in the T2K experiment}},  {\em IEEE Trans.Nucl.Sci.} {\bf
  57} (2010) 406--411.

\bibitem{MATACQ}
D.~Breton, E.~Delagnes, and M.~Houry, ``{Very high dynamic range and high
  sampling rate VME digitizing boards for physics experiments}.''
\newblock IEEE TRANSACTIONS ON NUCLEAR SCIENCE, VOL. 52, NO. 6, DECEMBER 2005.

\bibitem{DCalvet:Feminos}
D.~Calvet, ``{A New Versatile and Cost Effective Readout System for Small to
  Medium Scale Gaseous and Silicon Detectors}.''
\newblock To be presented at IEEE Nucl. Sci. Symposium, Oct. 27-Nov. 2, Seoul,
  Korea.

\bibitem{Iguaz:2011xj}
F.~Iguaz, T.~Dafni, E.~Ferrer-Ribas, J.~Galan, J.~Garcia, et~al., {\it {The
  Discrimination capabilities of Micromegas detectors at low energy}},  {\em
  Phys.Procedia} {\bf 37} (2012) 1079--1086,
  [\href{http://xxx.lanl.gov/abs/1110.2643}{{\tt arXiv:1110.2643}}].

\bibitem{GCantatore:SPSC2013}
G.~Cantatore, ``{SPSC status report October 2013 and request for 2014. In
  preparation, to be submitted by October 14th 2013}.''

\bibitem{Susana:mpgd2013}
S.~Cebri\'an, F.~Aznar, J.~F. Castel, et~al., ``{Assesment of material
  radiopurity for Rare Event experiments using Micromegas}.''
\newblock Proceedings of the 3rd International Conference on Micro Pattern
  Gaseous Detectors, MPGD2013, Zaragoza (Spain), 1-4 July 2013.

\bibitem{Dafni:2012fi}
T.~Dafni, S.~Aune, S.~Cebrian, G.~Fanourakis, E.~Ferrer-Ribas, et~al., {\it
  {Rare event searches based on micromegas detectors: The T-REX project}},
  {\em J. Phys. Conf. Ser.} {\bf 375} (2012) 022003.

\bibitem{Abbon:2007ug}
P.~Abbon et~al., {\it {The Micromegas detector of the CAST experiment}},  {\em
  New J. Phys.} {\bf 9} (2007) 170,
  [\href{http://xxx.lanl.gov/abs/physics/0702190}{{\tt physics/0702190}}].

\bibitem{ThPapaevangelou:SPSC2012}
T.~Papaevangelou, {\it {2012 Status report of the CAST Experiment and Running
  in 2013-2014}},  Tech. Rep. CERN-SPSC-2012-028, CERN, Geneva, Oct, 2012.

\bibitem{Iguaz:2012ur}
F.~Iguaz, E.~Ferrer-Ribas, A.~Giganon, and I.~Giomataris, {\it
  {Characterization of microbulk detectors in argon- and neon-based mixtures}},
   {\em JINST} {\bf 7} (2012) P04007,
  [\href{http://xxx.lanl.gov/abs/1201.3012}{{\tt arXiv:1201.3012}}].

\bibitem{Cebrian:2010nw}
S.~Cebri{\'a}n et~al., {\it {Micromegas readouts for double beta decay
  searches}},  {\em JCAP} {\bf 1010} (2010) 010,
  [\href{http://xxx.lanl.gov/abs/1009.1827}{{\tt arXiv:1009.1827}}].

\bibitem{theotesis}
T.~Vafeiadis, {\em Contribution to the search for solar axions in the CAST
  experiment}.
\newblock PhD thesis, Aristotle University of Thessaloniki, Thessaloniki
  (Greece), 2013.
\newblock CERN-THESIS-2012-349.

\bibitem{alfredotesis}
A.~Tom\'as, {\em Development of Time Projection Chambers with Micromegas for
  Rare Event Searches}.
\newblock PhD thesis, Universidad de Zaragoza, Zaragoza (Spain), 2013.
\newblock \emph{JINST} TH 001.

\bibitem{Thomas_SegmentedMesh}
T.~Papaevangelou, ``{Devolopment of Micromegas detectors for neutron
  time-of-flight measurements}.''
\newblock Proceedings of the 3rd International Conference on Micro Pattern
  Gaseous Detectors, MPGD2013, Zaragoza (Spain), 1-4 July 2013.

\end{thebibliography}\endgroup

\end{document}